\documentclass{article}
\begin{document}

\title{Oblique-Basis Calculations for $^{44}$Ti
\footnote{This work was supported in part by U.S. National
Science Foundation under grants (9970769 and 0140300) as well as a
Cooperative
Agreement (9720652) with matching from the
Louisiana Board of Regents Support
Fund.}}

\author{V. G. Gueorguiev, J. P. Draayer, W. E.
Ormand\footnote{Present address: Lawrence
Livermore National Laboratory,
Livermore, CA 94551}, and
C. W. Johnson\footnote{Present address: 
San Diego State University, San Diego, CA 92182}
\\ \it Department of Physics and Astronomy, 
\\ \it Louisiana State University,
\\ \it Baton Rouge, LA 70803}
\date{}
\maketitle

\begin{abstract}
The spectrum and wave functions of $^{44}$Ti are studied in
oblique-basis calculations using spherical and SU(3) shell-model states.
Although the results for $^{44}$Ti are not as good as those previously
reported for $^{24}$Mg, due primarily to the strong spin-orbit interaction
that generates significant splitting of the single-particle energies that
breaks the SU(3) symmetry, a more careful quantitative analysis shows that
the oblique-basis concept is still effective. In particular, a model space
that includes a few SU(3) irreducible representations, namely, the leading
irrep (12,0) and next to the leading irrep (10,1) including its spin $S=0$
and $1$ states, plus spherical shell-model configurations (SSMC) that have
at least two valence nucleons confined to the $f_{7/2}$ orbit -- the SM(2)
states, provide results that are compatible with SSMC with at least one
valence nucleon confined to the $f_{7/2}$ orbit -- the SM(3) states.
\end{abstract}

\textit{Introduction.} In a previous study we demonstrated the feasibility
of the oblique-basis calculations.\cite{VGG 24MgObliqueCalculations}
The successful description of $^{24}$Mg followed from the comparable
importance of single-particle excitations, described by spherical
shell-model configurations (SSMC), and collective excitations, described by
the SU(3) shell model.  An important element of the success is that SU(3)
is a good symmetry in $sd$-shell nuclei.\cite{Elliott's SU(3) model} For
the lower $pf$-shell nuclei, there is strong breaking of the SU(3) symmetry
induced by the spin-orbit interaction.\cite{VGG SU(3)andLSinPF-ShellNuclei}
Therefore, it is anticipated that adding the leading and next to the
leading SU(3) irreps may not be sufficient in lower $pf$-shell.

Here we discuss oblique-basis type calculations for $^{44}$Ti using the KB3
interaction.\cite{KB3 interaction} We confirm that the spherical shell
model (SSM) provides a significant part of the low-energy wave functions
within a relatively small number of SSMC while a pure SU(3)
shell-model with only few SU(3) irreps is unsatisfactory. This is the
opposite of the situation in the lower $sd$-shell. Since the SSM yields
relatively good results for SM(2), combining the two basis sets yields
even better results with only a very small increase in the overall size of
the model space. In particular, results in a SM(2)+SU(3) model space
(47.7\% + 2.1\% of the full $pf$-shell space) are comparable with SM(3)
results (84\%). Therefore, as for the $sd$-shell, combining a few SU(3)
irreps with SM(2) configurations yields excellent results, such as correct
spectral structure, lower ground-state energy, and improved structure of
the wave functions. However, in the lower $sd$-shell SU(3) is dominant and
SSM is recessive (but important) and in the lower $pf$-shell one finds the
opposite, that is, SSM is dominant and SU(3) is recessive (but important).

\textit{Model Space.}
$^{44}$Ti consists of 2 valence protons and 2 valence neutrons in the
$pf$-shell. The SU(3) basis includes the leading irrep (12,0) with
$M_{J}=0$ dimensionality 7, and the next to the leading irrep (10,1). The
(10,1) occurs three times, once with $S=0$ (dimensionality 11) and twice
with $S=1$ (dimensionality $2\times 33=66$). All three (10,1) irreps have
a total dimensionality of 77. The (12,0)\&(10,1) case has a total
dimensionality of 84 and is denoted by \&(10,1). In Table \ref{TableTi44}
we summarize the dimensionalities. As in the case of $^{24}$Mg, there
are linearly dependent vectors within the oblique bases sets. For example,
there is one redundant vector in the SM(2)+(12,0) space, two in
SM(3)+(12,0) and SM(1)+(12,0)\&(10,1) spaces, twelve in
SM(2)+(12,0)\&(10,1) space, and thirty-three in the SM(3)+(12,0)\&(10,1)
space. Each linearly dependent vector is handled as in the previous
case.\cite{VGG 24MgObliqueCalculations}

\begin{table}[tbp]
\caption{Labels and M$_{J}$=0 dimensions for various $^{44}$Ti
calculations. The leading SU(3) irrep is (12,0); \&(10,1) implies that the
(10,1) irreps are included along with the leading irrep. SM(n) is a
spherical shell-model basis with n valence particles anywhere within the
full $pf$-shell; the remaining particles being confined to the $f_{7/2}$.}
\begin{tabular}{|lrrrrrrr|}
\hline
Model space & (12,0) & \&(10,1) & SM(0) & SM(1) & SM(2) & SM(3) & FULL\\
\hline
dimension & 7 & 84 & 72 & 580 & 1908 & 3360 & 4000 \\
dimension \% & 0.18 & 2.1 & 1.8 & 14.5 & 47.7 & 84 & 100 \\
\hline
\end{tabular}\label{TableTi44}
\end{table}

\textit{Ground-state Energy.} The oblique-basis calculation of the
ground-state energy for $^{44}$Ti does not look as impressive as for
$^{24}$Mg. The calculated ground-state energy for the SM(1)+(12,0)\&(10,1)
space is $0.85$ MeV below the calculated energy for the SM(1) space. Adding
the two SU(3) irreps to the SM(1) basis increases the size of the space
from 14.5\% to 16.6\% of the full space. This is a 2.1\% increase, while
going from the SM(1) to SM(2) involves an increase of 33.2\%. For SM(2),
the ground-state energy is $2.2$ MeV lower than the SM(1) result. However,
adding the SU(3) irreps to the SM(2) basis gives ground-state energy of
$-13.76$ MeV which is compatible to the pure SM(3) result of $-13.74$ MeV.
Therefore, adding the SU(3) to the SM(2) increases the model space
from 47.7\% to 49.8\% and gives results that are slightly better than the
SM(3) which is 84\% of the full space.

\textit{Low-lying Energy Spectrum.} In $^{24}$Mg the position of the K=2
band head is correct for the SU(3)-type calculations but not for the
low-dimensional SM(n) calculations.\cite{VGG 24MgObliqueCalculations} In
$^{44}$Ti it is the opposite, that is, the SM(n)-type calculations
reproduce the position of the K=2 band head while SU(3)-type calculations
cannot. Furthermore, the low-energy levels for the SU(3) case are higher 
than for the SM(n) case. Nonetheless, the spectral structure in the
oblique-basis calculation is good and the SM(2)+(12,0)\&(10,1) spectrum
($\approx$50\% of the full space) is comparable with the SM(3) result
(84\%).

\textit{Overlaps with Exact States.} The overlap of SU(3)-type calculated
eigenstates with the exact (full shell-model) results are not as large as
in the $sd$-shell, often less than 40\%, but the SM(n) results are
considerably better with SM(2)-type calculations yielding an 80\% overlap
with the exact states while the results for SM(3) show overlaps greater
than 97\%, which is consistent with the fact that SM(3) covers 84\% of the
full space. On the other hand, SM(2)+(12,0)\&(10,1)-type calculations yield
results that are as good as those for SM(3) in only about 50\% of the
full-space and  SM(1)+(12,0)\&(10,1) overlaps are often bigger than the
SM(2) overlaps.

\textit{Conclusion.} For $^{44}$Ti, combining a few SU(3) irreps with SM(2)
configurations increases the model space only by a small ($\approx$2.3\%)
amount but results in better overall results: a lower ground-state energy,
the correct spectral structure (particularly the position of K=$2^+$ band
head), and wave functions with a larger overlap with the exact results.
The oblique-bases SM(2)+(12,0)\&(10,1) results for $^{44}$Ti
($\approx$50\%) yields results that are comparable with the SM(3) results
($\approx$84\%). In short, the oblique-basis scheme works well for
$^{44}$Ti, only in this case, in contrast with the previous results for
$^{24}$Mg where SU(3) was found to be dominant and SSM recessive, in the
lower $pf$-shell SSM is dominant and SU(3) recessive.

\end{document}